# A Cluster Based Replication Architecture for Load Balancing in Peer-to-Peer Content Distribution

**S.Ayyasamy[1] and S.N. Sivanandam[2]**

[1]Asst. Professor, Department of Information Technology,
Tamilnadu College of Engineering Coimbatore-641 659, Tamil Nadu, INDIA.
Email: ayyasamyphd@gmail.com

[2]Professor and Head, Department of Computer Science and Engineering,
PSG College of Technology, Peelamedu, Coimbatore-641 004, Tamil Nadu, INDIA.

## *Abstract*

*In P2P systems, large volumes of data are declustered naturally across a large number of peers. But it is very difficult to control the initial data distribution because every user has the freedom to share any data with other users. The system scalability can be improved by distributing the load across multiple servers which is proposed by replication. The large scale content distribution systems were improved broadly using the replication techniques. The demanded contents can be brought closer to the clients by multiplying the source of information geographically, which in turn reduce both the access latency and the network traffic. In addition to this, due to the intrinsic dynamism of the P2P environment, static data distribution cannot be expected to guarantee good load balancing. If the hot peers become bottleneck, it leads to increased user response time and significant performance degradation of the system. Hence an effective load balancing mechanism is necessary in such cases and it can be attained efficiently by intelligent data replication.*

*In this paper, we propose a cluster based replication architecture for load-balancing in peer-to-peer content distribution systems. In addition to an intelligent replica placement technique, it also consists of an effective load balancing technique. In the intelligent replica placement technique, peers are grouped into strong and weak clusters based on their weight vector which comprises available capacity, CPU speed, access latency and memory size. In order to achieve complete load balancing across the system, an intra-cluster and inter-cluster load balancing algorithms are proposed. We are able to show that our proposed architecture attains less latency and better throughput with reduced bandwidth usage, through the simulation results.*

## 1. Introduction

### P2P Overlay Networks

To share the computer resources like content, storage, CPU cycles directly without using an intermediate system or a centralized server, distributed computer architecture, called "peer-to-peer" are designed. They are distinguished by their failure adaptation capabilities and maintenance of acceptable connectivity and performance [1]. Significant research attention has been applied to Content distribution, which is an important peer-to-peer application on the Internet. By allowing personal computers to work as a distributed storage medium, they normally contribute, search and obtain digital content.

Overlays are flexible and deployable approaches that allow users to perform distributed operations without modifying the underlying physical network. Peer-to-peer (P2P) overlay systems have been proposed to address a variety of problems and enable new applications. The attraction of these systems, when compared to client/server frameworks, is in their robustness, reliability and cost efficiency.

Unlike traditional distributed computing, P2P networks aggregate large number of computers and possibly mobile or handheld devices, which join and leave the network frequently. Nodes in a P2P network, called peers, play a variety of roles in their interaction with other peers. When accessing information, they are clients. When serving information to other peers, they are servers. When forwarding information for other peers, they are routers. This new breed of systems creates application-level virtual networks with their own overlay topology and routing protocols.

To search for data or resources, messages are sent over multiple hops from one peer to another with each peer responding to queries for information it has stored locally. Structured P2P overlays implement a distributed hash table data structure in which every data item can be located within a small number of hops at the expense of keeping some state information locally at the nodes.

## Replica Placement for QoS-Aware Content Distribution

Replication techniques are widely employed to improve the availability of data, enhancing performance of query latency and load balancing, in content distribution systems. We can geographically multiply the source of information by distributing multiple copies of data in the network. By forwarding each query to its nearest copy, the query search latency can be effectively reduced.

The ability to improve system scalability through distributing the load across multiple servers [2] is also offered by replication. If a replica of the requested object (e.g., a web page or an image) is kept in its nearer proximity then the clients would feel low access latency. Depending on the position of the replicas, the effectiveness of replication tends to a large extent.

The centralized servers become a bottleneck as the requirement of the information increases. The performance problem is managed by the content providers, system administrators or end users by themselves through delivering replicas of web content to machines, spread throughout the network. The load on the central server [3] is reduced by replicas through responding to the local client requests. The load which is delivered to the cooperate nodes includes:

- Communication bandwidth, for sending the data to the requesting content,
- Storage used for hosting the replica and
- CPU resources for query processing.

The problem of deciding how many replicas is to be delivered to each file and its location is given by the Replica management to this circumstances. To handle more requirements for each file, enough replicas should be present. Servers become overloaded and clients observe lower performance by having only few replicas. On the other hand the waste bandwidth of extra replicas and the storage which could be reassigned to the other files, and also the money spent to rent, power and also for host machine cooling.

### Load Balancing

In P2P systems, large volumes of data are declustered naturally across a large number of peers. But it is very difficult to control the initial data distribution because every user has the freedom to share any data with other users. In addition to this, due to the intrinsic dynamism of the P2P environment, static data distribution cannot be expected to guarantee good load balancing. In some of the hot peers, the number of disk accesses is unequal because of changing the popularities of various data items and skewed query patterns. Therefore this causes severe load imbalance throughout the system. If the hot peers become bottleneck, it leads to increased user response time and significant performance degradation of the system. Hence the load balancing mechanism is necessary in such cases and it can be attained efficiently by online data migration/replication.

In this paper, we propose a cluster based replication architecture for load-balancing in peer-to-peer content distribution systems. It contains an intelligent replica placement algorithm with an effective load balancing technique. This paper is an extension of our previous work [18].

This paper is organized as follows. Section 2 gives the detailed related work done. Section 3 presents the system model and algorithm overview for the proposed architecture. Section 4 presents the intelligent replica placement algorithm, followed by the searching technique. Section 5 describes the load balancing technique in detail. Section 6 gives the experimental results and section 7 concludes the paper.

## 2. Related Works

Most of the research efforts to improve the performance of Gnutella-like P2P systems can be classified into two categories:

1) P2P search and routing algorithms and
2) P2P overlay topologies.

Most of the proposed routing or search algorithms in the first category, disregard the natural peer heterogeneity present in most P2P systems and more importantly the potential performance hurdle caused by the randomly constructed overlay topology.

B. Mortazavi_ and G. Kesidis [4] have provided a survey of reputation systems. Based on a reputation framework, they have designed a game in which users play to maximize the received files from the system. For this, the users adjust their cooperation level, there by obtaining a good reputation as a result.

Brighten Godfrey et al [5] have proposed an algorithm for load balancing in heterogeneous and dynamic P2P systems. Their simulation results shows that in the face of rapid arrivals and departures of objects of widely varying load, their algorithm achieves load balancing for system utilizations as high as 90% while moving only about 8% of the load that arrives into the system. Similarly, in a dynamic system where nodes arrive and depart, their algorithm moves less than 60% of the load the underlying DHT moves due to node arrivals and departures. Finally, they have shown that their distributed algorithm performs only negligibly worse than a similar centralized algorithm and that node heterogeneity helps, not hurts, which is the scalability of their algorithm.

Kalman Graffi et al [6] have proposed a DHT-based information gathering and analyzing architecture which controls the streaming request assignment in the system and thoroughly evaluate it in comparison to a distributed stateless strategy. They evaluated the impact of the key parameters in the allocation function which considers the capabilities of the nodes and their contribution to the system. Identifying the quality-bandwidth tradeoffs of the information gathering system, they illustrate that with their proposed system a 53% better load balancing can be reached and the efficiency of the system is significantly improved.

Paraskevi Raftopoulou and Euripides G.M. Petrakis have presented iCluster, a self-organizing peer-to-peer overlay network for supporting full-fledged information retrieval in a dynamic environment. They defined the criteria for peer similarity and peer selection, and also presented the protocols for organizing the peers into clusters and for searching within the clustered organization of peers [7].

Unfortunately, most existing work on replica placement has focused on optimizing an average performance measure of the entire client community such as the mean access latency [8], [9]. While an average performance measure may be important from the system's point of view, it does not differentiate the likely diverse performance requirements of the individuals. So far, to the best of our knowledge, there has been no study on QoS-aware replica placement.

Carvalho, N. Araujo, F. Rodrigues. L, have presented the IndiQoS architecture, a scalable QoS-aware publish-subscribe system with QoS-aware publications and subscriptions that preserves the decoupling which makes the publish-subscribe model so appealing. To support such model, the proposed architecture IndiQoS includes a decentralized message-broker based on a DHT that leverages on underlying network-level QoS reservation mechanisms [10].

Guillaume Pierre and Maarten van Steen have presented Globule, a collaborative content delivery network. The Proposed network was composed of Web servers that cooperate across a wide-area network to provide performance and availability guarantees to the sites they host [12].

David Novak [14] suggested a new general solution of the load-balancing problem in P2P Data Networks, which is especially suitable for systems with time consuming search operations. The proposed framework analyzes the source of the load precisely to choose right balancing action.

The scalability and performance of DHTs is strongly based on an equal distribution of data across participating nodes. Because this concept is based on hash functions, one assumes that the content is distributed nearly evenly across all DHT-nodes. Nonetheless, most DHTs show difficulties in load balancing as we will point out in this paper. To ensure the major advantages of DHTs – namely scalability, flexibility and resilience Simon Rieche et al [15] have discussed three approaches of load balancing and compare them corresponding to simulation results.

Theoni Pitoura et al [16] have presented Hot-RoD, a DHT-based architecture that deals effectively with this combined problem through the use of a novel locality-preserving hash function, and a tunable data replication mechanism which allows trading off replication costs for fair load distribution. Their detailed experimentation study shows strong gains in both range query processing efficiency and data-access load balancing, with low replication overhead.

Ananth Rao et al [17] have addressed the problem of load balancing in P2P systems. They explored the space of designing load-balancing algorithms which uses the notion of "virtual servers". They have presented three schemes that differ primarily in the amount of information used to decide how to re-arrange load. Their simulation result shows that even the simplest scheme is able to balance the load within 80% of the optimal value, while the most complex scheme is able to balance the load within 95% of the optimal value.

Song Fu et al [18] characterized the behaviors of randomized search schemes in the general P2P environment. They extended the supermarket model by investigating the impact of node heterogeneity and churn to the load distribution in P2P networks. They have proved that by using d-way random choices schemes, the length of the longest queue in P2P systems with heterogeneous nodal capacity and node churn for $d \geq 2$ is clog logn/logd + O(1) with high probability, where c is a constant.

## 3. System Model and Algorithm Overview

### Algorithm Overview

In our QOS aware topology, nodes are grouped into strong and weak clusters based on their weight vector which comprises the following parameters:

- Available capacity
- CPU speed
- Memory size
- Access Latency

In the replica placement algorithm, we classify the content as Class I and Class II, based on their access patterns. (i.e.) The most frequently accessed contents are ranked as Class I and the less frequently accessed contents as Class II. Then more copies of Class I content are replicated in strong clusters (having high weight values). Routing is performed hierarchically by broadcasting the query only to the strong clusters. Thus our proposed architecture achieves Low bandwidth Consumption, Reduced Latency, Reduced Maintenance Cost, Strong Connectivity and Query Coverage.

### System Model

Let us consider a collection of N server nodes which form a peer to peer (P2P) overlay network. In addition to being part of the overlay, each node functions as a server responding to requests (queries) which come from clients outside of the overlay network. An example could be that each node is a web server with the overlay linking the servers and clients being web browsers on remote machines requesting content from the servers.

We assume each node always stores one copy of its own content item which it serves to clients and that it has additional storage space to store k replicated content items from other nodes which it can also serve [3]. The object is associated with an authoritative *origin server* (OS) in the network where the content provider makes the updates to the object. The object copy located at the origin server is called the *origin copy* and an object copy at any remaining server is called a *replica*.

# 4. Intelligent Replica Placement Algorithm

## Clustering the Nodes

*For* each node $N_i$, i = 1,2.......n, let

$BW_i$ - Available Bandwidth

$SP_i$ - CPU Speed

$AL_i$ - Access Latency

$MZ_i$ - Memory Size

1. The weight of the node Ni can be calculated as

$$W_i = (BW_i + SP_i + MZ_i) / AL_i$$

2. Form the vector $W = \{S_i, W_i\}$, which denotes the node ids and their corresponding weight values, sorted on the descending order.

3. Let {Sk} denote the set of strong cluster nodes $(0 <= k < n)$, which satisfies the following condition $W_k \geq \beta$, where $\beta$ is the minimum threshold value for the weight.

4. Then the set {Wj} = {Ni} – {Sk}, denote the set of weak cluster nodes $(0 <= j < n)$, which satisfies the condition $W_k < \beta$

## Replica Placement

Let QS be the query server which registers the query of each client. The query server stores the cluster information of each node along with the node id as "S" or "W" for strong and weak clusters, respectively.

1. At time Tk, let m clients generates query requests {Qm} of the form q {nid, ckwd}, where nid is the node id of the client and ckwd is the keyword of the content to be retrieved.

2. The queires {Qm} are registered in the query server QS.

3. The requested content of the queries are classified and categorized as class1 or class2, depending on the access frequencies.
   (i.e.) A query Qj, j<m, is considered to be class1
   If n (Qj) >= Amin
   and class2,
   If n (Qj) < Amin
   Where n (Qj) is the no. of access of the content pattern for the given query and Amin is the minimum access threshold value.

4. Then the query server QS assigns the class1 contents to the strong cluster nodes and class2 contents to the weak cluster nodes.

5. After the assignment, QS transmit these replication pattern information to the origin server OS.

6. OS performs the replication placement, according to the pattern information obtained from QS. The weight value Wi of each node is stored along with the content.

7. OS then broadcasts the replication information to the respective clients in the following format

{Nid, Clid ("S" or "W"), c1, c2 …}

Where Nid is the node id, Clid is the cluster id and c1, c2… are content database ids.

# 5. Load Balancing Through Replication

In this section, we present the intra-cluster and inter-cluster load balancing through the replication of data. Here, balancing the load within a particular cluster is called as intra-cluster load balancing and balancing the load among the clusters is called as inter-cluster load balancing. This is done in order to achieve complete load balancing across the system.

## Replication Constraints

Load balancing can be attained through data replication by transferring hot data from heavily loaded peers to lightly loaded peers. Since the search is entirely distributed, a particular replica of a specified data item Di is accessed for large number of times rather than using many replicas. Thus it does not provide absolute guarantee of load balancing. Even though there is a cost of disk space, replication increases the data availability. Since the data which was hot previously may become cold subsequently, a periodic cleanup of the replicas is necessary. Therefore this shows that the replicas are no longer needed. In addition to this, issues of replicating large data items are need to be examined. Basically, our main objective is to make sure that the replication executed for short-term benefit does not cause long-term degradation in system performance by causing unwanted wastage of valuable disk space at the peers.

We propose that the run-time decision for both intra-cluster and inter-cluster which involves replication should be made as follows: Every cluster leader observes it peer's availability over a period of time. The hot data should be replicated for availability reasons if the probability of a peer $P_1$ leaving the system is high. It is to be noticed that the replication will be done only if it is subjected to disk space constraints at the destination peers. In addition to this, if the disk capacities of the peers are larger than that of the size of the large data items then the large data items shall be replicated.

Each peer $P_i$ maintains the set of data items D replicated at it. However, to determine the data items which are still hot, $P_i$ checks periodically the number of accesses $N_k$ for the last time interval on each item in D. The items for which $N_k$ is less than a predefined threshold $\alpha$ are deleted since those items may not be hot anymore. Thus, it eliminates the need for replication.

Consider the hot data items are numbered as $H_1, H_2, H_3 \ldots H_m$ ($H_1$ is the hottest element) in a cluster. The original copy of these replicas is stored at the peers in which started the replication and it shows that the original data item is not deleted. To provide the system scalability over time, it is important to delete the replicas periodically, which yields more disk space.

## Intra-Cluster Load Balancing

In case of intra-cluster load balancing, some of the decisions are critical to system performances regarding when to trigger the load balancing mechanism, hotspot detection and the amount of data to be replicated.

In this approach, the workload statistics of each peer is sent to its cluster leader periodically. Load balancing is started when the cluster leader detects a load imbalance in the cluster.

In the conventional domains, intra-cluster load balancing are researched perfectly. But for P2P systems, we should include the changing available disk capacities of the peers.

The cluster leader $CL_i$ receives the information periodically regarding the loads $W_i$ and available disk space Si of the peers. Based on Wi the cluster leader CLi creates a sorted list $l_i$ of the peers such that the first element of the list is the heavily loaded peer. Let us consider that there are n elements in the list. Among the last [n/2] peers in the list, the peers whose corresponding values of Si which are less than a pre-specified threshold $S_{th}$ are deleted. Now the load balancing is achieved by replicating the hot data H from the first peer in the list to the last peer and the second peer to the second last peer and so on. If the load difference between the peers exceeds a pre-specified threshold β, then the data will be replicated. We can monitor that, only for a particular periodic time intervals $CL_i$ checks for the load imbalance and not whenever any peer joins/leaves the system. $CL_i$ corrects the load imbalances which are caused by some peers while joining/leaving the system. These are done only at the next periodic time intervals. We trust that while performing load balancing every time, a peer joins/leaves will results in disastrous condition because peers may join/leave the system frequently. The above steps are summarized in the following algorithm.

### Algorithm –Intra Cluster Load Balancing

1. For each $\{CL_i\}^{k}_{i=1}$
2. For each member $\{Pj\}^{n}_{j=1}$ of CLi
3. $P_{j,i}$ send Wj, i and $S_{j,i}$ to $CL_i$
4. $CL_i$ add $P_{j,i}$ to the list $\{l_i\}$
5. End For
6. $CL_i$ sort $l_i$ such that $W_{j,i} > W_{j-1,i} > W_{j-2,i}$…..
7. For each $\{l_j\}^{n}_{j=n/2}$
8. If $S_{j,i} < S_{th}$ then
9. Delete the element $P_{j,i}$
10. End if
11. End For
12 If $Wa - W_b > \beta$ for any a, b < n, then
13. Move H1 ($N_1$) into $N_n$.
H2 ($N_2$) into $N_{n-1}$ and so on.
14. End if
15. End For
16. End

## Inter Cluster Load Balancing

Inter cluster load balancing is necessary in order to prevent load imbalance among the clusters. We propose that such load balancing should be carried out between the neighboring clusters through cooperation between the cluster leaders. This is because moving data to distant clusters may obtain high communication overhead to align the movement.

The load information is exchanged from the cluster leaders with their neighboring cluster leaders periodically. The cluster leader $CL_k$ checks whether its load exceeds the average load of the set $\{CL_i\}$ of its neighboring cluster leaders by more than 10% of the average load. If it exceeds, then it determines the hot data items which should be moved. It sends a message about the disk space requirements of each data items to each cluster leader in $\{CL_i\}$ to transfer some part of its load to them. The leaders in $\{CL_i\}$ check the available disk space in each of their cluster members. They send a message to $CL_k$ about the total loads and their available disk space if the space limits are satisfied. Therefore $CL_k$ arranges the leaders which are ready in $\{CL_i\}$ to the List $l_k$ so that the first element of the List $l_k$ is the least loaded leader.

We assume that r denotes the number of willing peers in $\{CL_i\}$ and m denotes the number of hot data items. If $r < m$, then $H_1$ is assigned to the first element of $l_k$ and $H_2$ is assigned to the second element and so on in a round-robin fashion. This is done until all the hot items have been assigned. Suppose if $r \geq m$, then the assignment of hot data to elements of $l_k$ is performed in the same way as above. But in this case some elements of $l_k$ will not receive any hot data.

Once the hot data arrived at the cluster leader $CL_i$, then the leader creates a sorted list $l_i$ in descending order of load of its peers. Then using the intra-cluster load balancing, the cluster leader assigns the hot data to the elements of $l_i$.The above steps are summarized in the following algorithm

**Algorithm- Inter Cluster Load Balancing**

1. Consider a cluster leader $CL_k$.
2. $CL_k$ exchanges $\{Wi\}$ with $\{CL_i\}^n_{i=1}$
3. If $(Wk - W_{avg}) > (W_{avg} * 0.10)$ Then
   ($W_{avg}$ is the average load of $\{CL_i\}^n_{i=1}$ and
   Wk is the load of $CL_k$ )
4. For each member $\{Pj\}^n_{j=1}$ of $CL_k$
5.     $CL_k$ send $S_j$ to $\{CL_i\}^n_{i=1}$
6. End For
7. For each $\{CL_i\}^n_{i=1}$
8.     For each member $\{P_v\}^n_{v=1}$ of CLi
9.         If $S_{v,i} > Min(\{S_j\}^n_{j=1})$ Then
10.             Send $S_{v,i}$ and $W_{v,i}$ to $CL_i$
11.         End if
12     End For
13.     $CL_i$ sends $\sum W_{v,i}$ and $\sum S_{v,i}$ to $CL_k$
14.     $CL_k$ add $CL_i$ to the list $\{l_k\}$
15. End For
16. $CL_k$ sort $l_k$ such that $\sum W_{v,i} < \sum W_{v+1,i} < \sum W_{v+2,i} \ldots$
17. If r < m Then

```
    For each Hi of CL_k
        Move H1 into CL1,
            H2 into CL2 and so on.
    End For
End if
Apply Intra-Cluster load balancing to {CL_i} n i=1
End if
End.
```

# 6. Experimental Results

## Simulation Setup

This section deals with the experimental performance evaluation of our algorithms through simulations. In order to test our protocol, the NS2 simulator is used. NS2 is a general-purpose simulation tool that provides discrete event simulation of user defined networks.

We have used the Bit Torrent packet-level simulator for P2P networks [13]. A network topology is only used for the packet-level simulator. Based on the assumption that the bottleneck of the network is at the access links of the users and not at the routers, we use a simplified topology in our simulations. We model the network with the help of access and overlay links. Each peer is connected with an asymmetric link to its access router. All access routers are connected directly to each other modeling only an overlay link. This enables us to simulate different upload and download capacities as well as different end-to-end (e2e) delays between different peers.

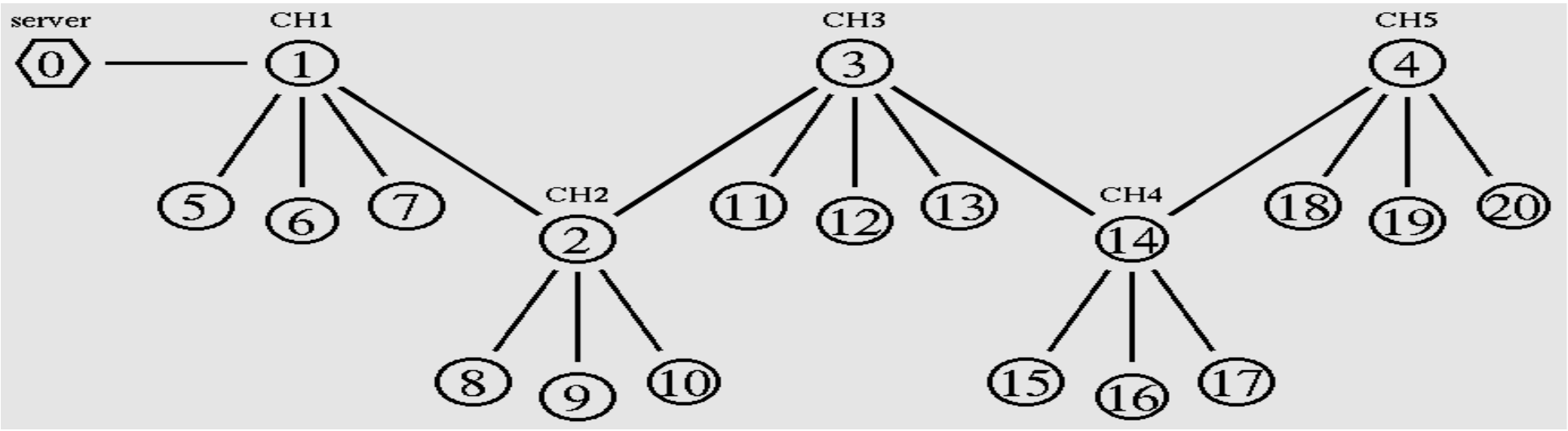

Fig. 1: Topology of P2P overlay network

## Simulation Results

We have simulated our Cluster Based Replication architecture with load balancing (WithLB) and without load balancing (WithoutLB) and measure the throughput, delay and packet loss.

### Based On Load

In our initial experiment, the load of the requested content is varied from 250bytes to 2000bytes. The response delay and received throughput are measured. In Figure 2, we can see that, when the load increases, the delay also increases. It is evident that the delay of LB is significantly less than the delay of WithoutLB. Figure 3 shows the aggregated throughput of all the client nodes which obtained their respective share of data. From the figure we can see that the LB has more throughput than WithoutLB.

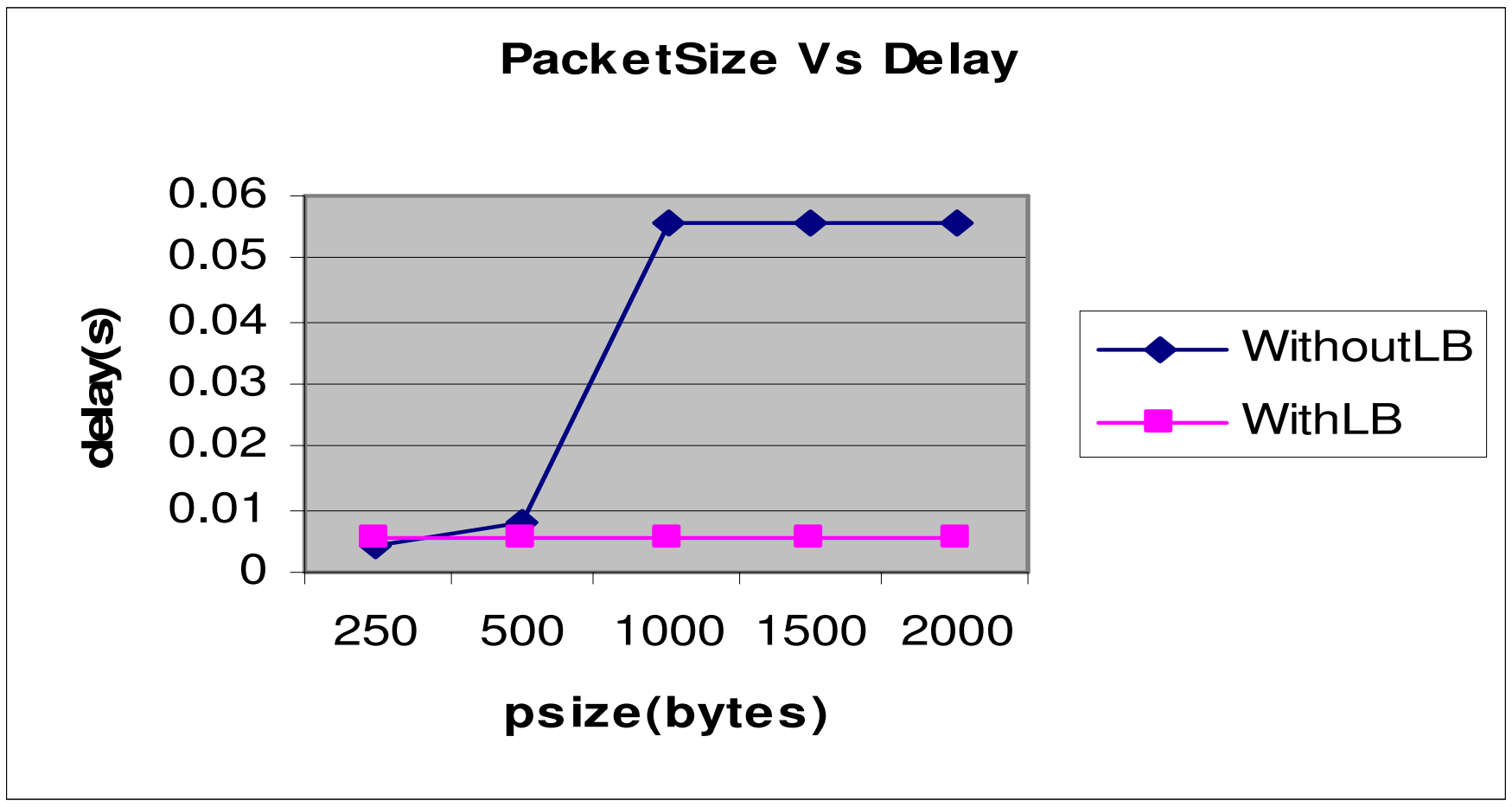

Fig. 2: Load Vs Delay

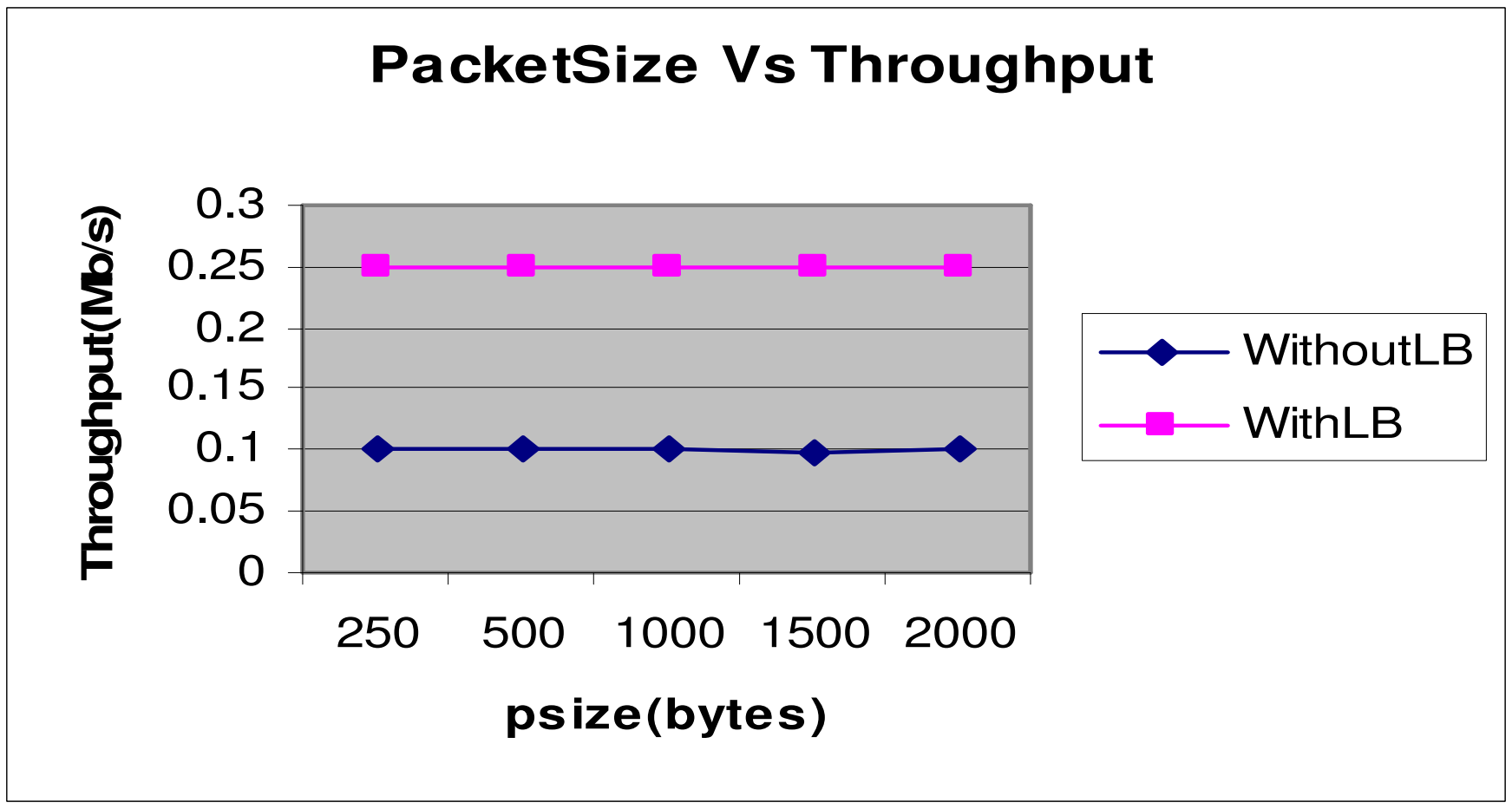

Fig. 3: Load Vs Throughput

**Based On Rate**

In our second experiment, the query sending rate is varied from 250Kb to 1Mb. The response delay and received throughput are measured. In Figure 4, we can observe that, when the rate increases, the delay remains almost constant for WithoutLB but decreases in the case of LB. From the figure, it can be seen that the delay of LB is significantly less than the delay of WithoutLB.

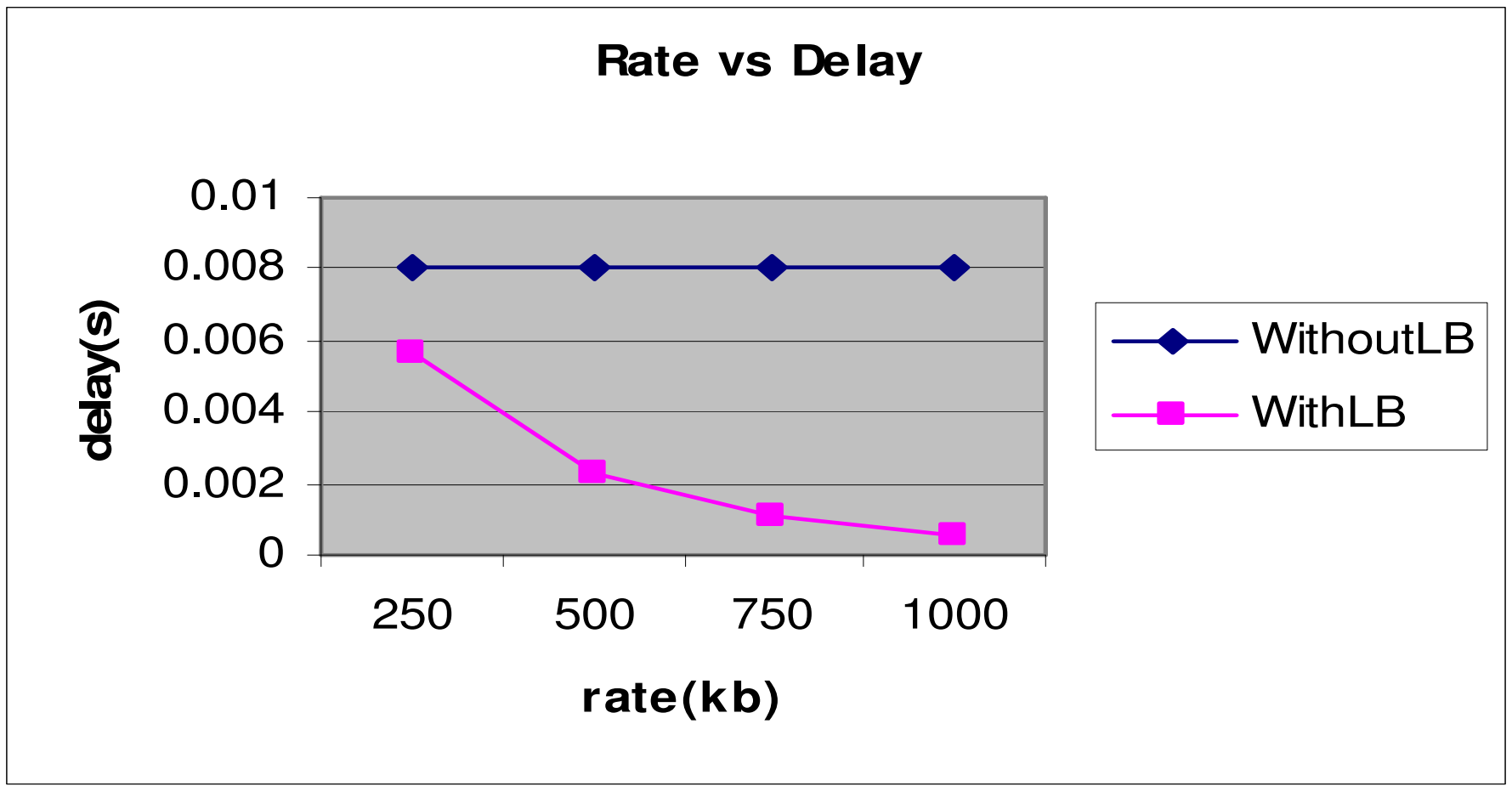

Fig. 4: Rate Vs Delay

In Figure 5, the throughput against rate is shown. From the figure, we can see that the throughput of LB is more when compared to WithoutLB, and increases when rate increases.

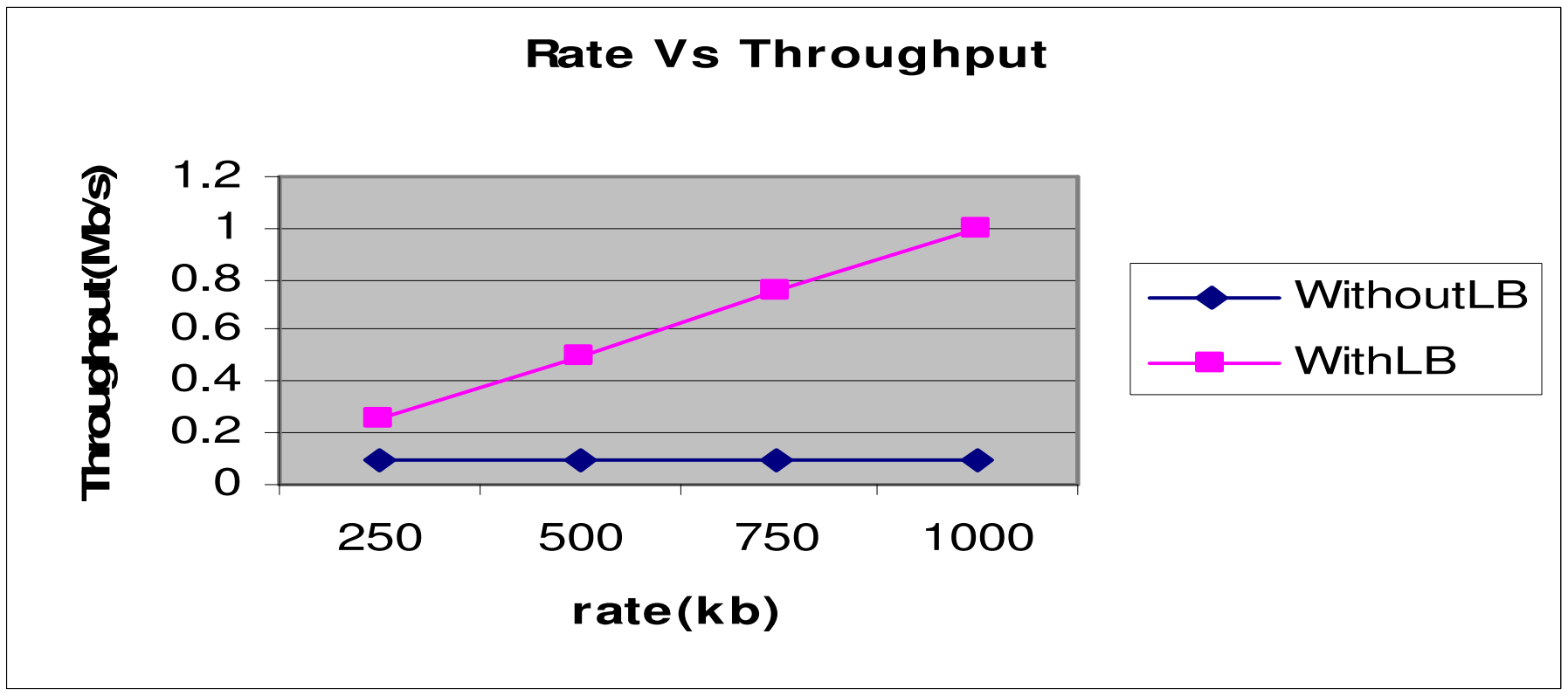

Fig. 5: Rate Vs Throughput

**Based on Simulation Time**

In our last experiment, the simulation time is varied from 10 to 20 seconds. The response delay, packets lost and received throughput are measured.

In Figure 6, the throughput against time is shown. From the figure, we can see that the throughput of LB is more when compared to WithoutLB, and remains constant when time increases.

In Figure 7, we can see that the delay of LB is significantly less than the delay of WithoutLB.

The number of packets lost is shown in Figure 8. As the time increases, the packet lost also increases in the case of WithoutLB. For LB, ultimately there is no packet loss.

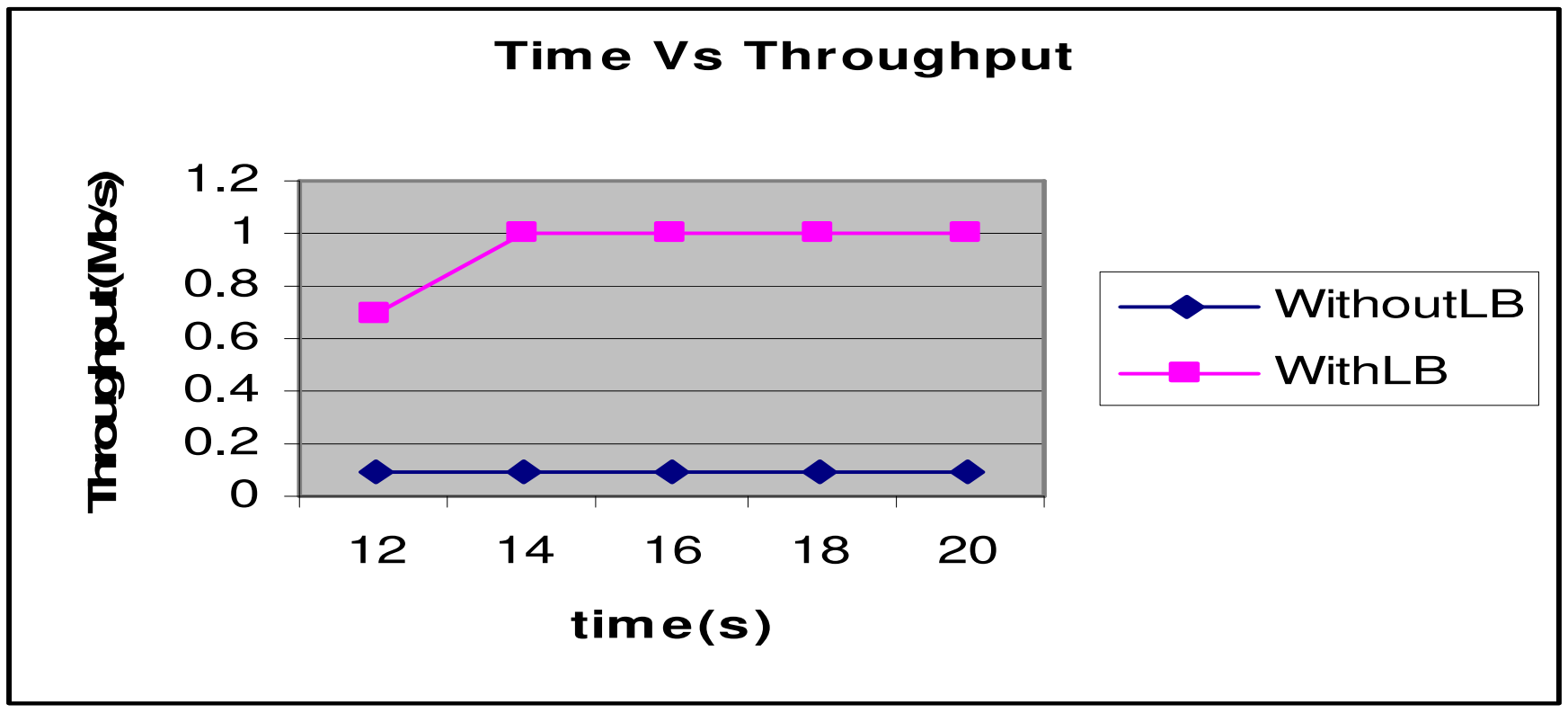

Fig. 6: Time Vs Throughput

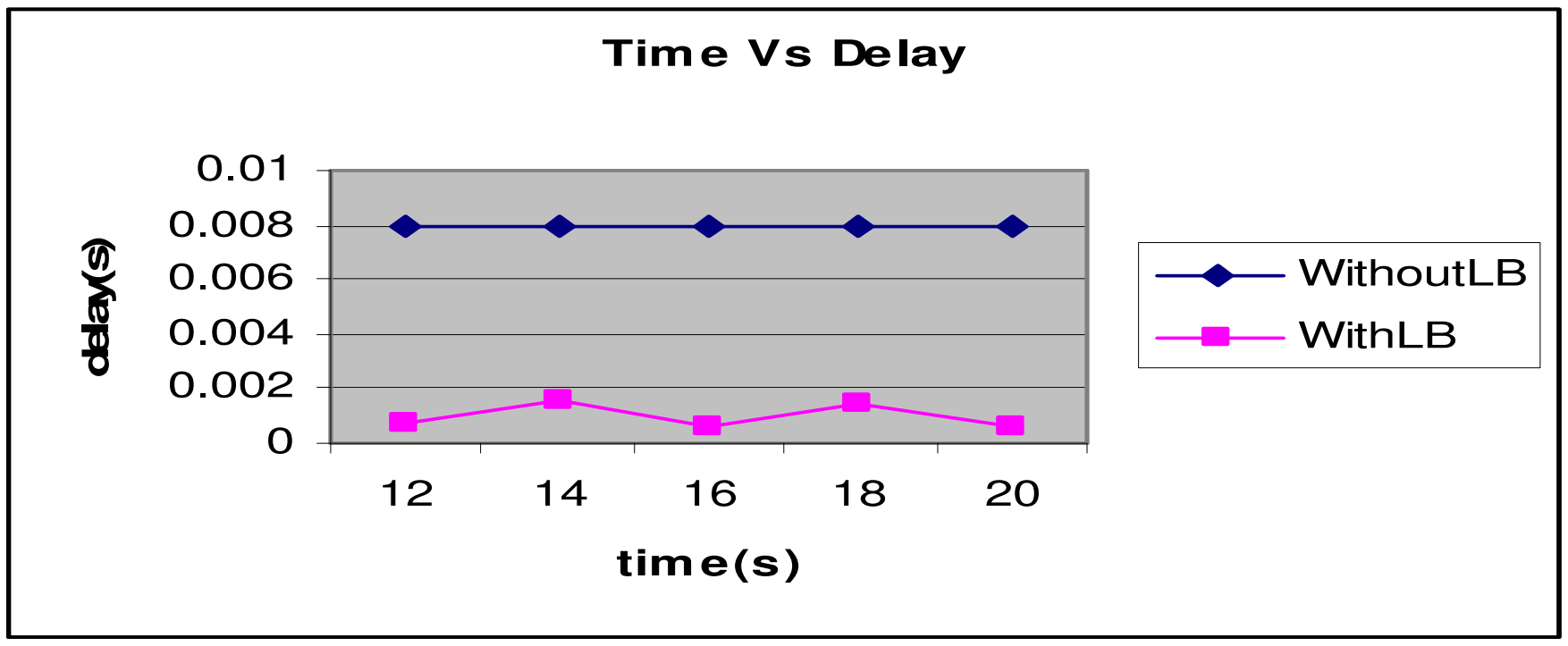

Fig. 7: Time Vs Delay

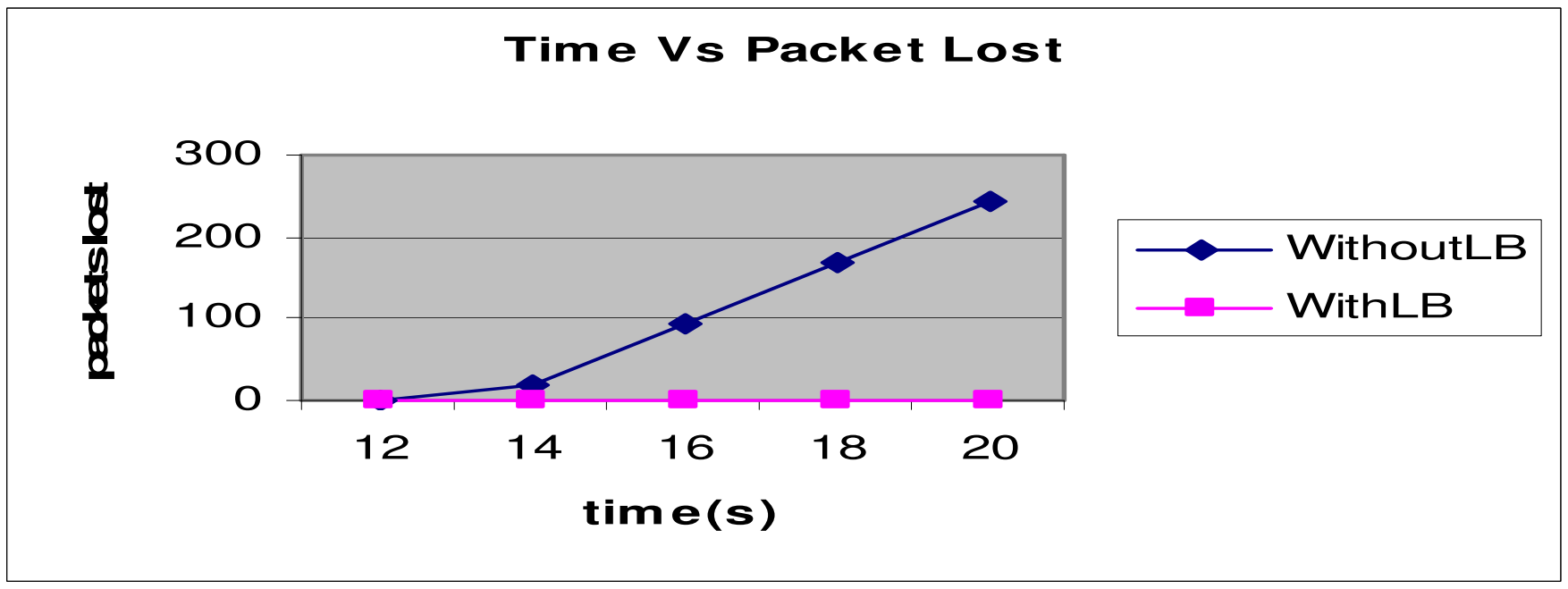

Fig.8: Time Vs Packet Lost

## 7. Conclusion

In this paper, we have proposed a cluster based replication architecture for load-balancing in peer-to-peer content distribution systems. Based on the weight vector which includes available capacity, CPU speed, and memory size and access latency the nodes are classified into strong and weak clusters. Based on the access pattern the content is classified into class I or class II by the replica management algorithm. Then class I contents are replicated into strong groups for more copies. Routing is performed only to the strong clusters through broadcasting the query hierarchically. In addition to an intelligent replica placement technique, it also consists of an effective load balancing technique. In the intelligent replica placement technique, peers are grouped into strong and weak clusters based on their weight vector which comprises available capacity, CPU speed, memory size and access latency. In order to achieve complete load balancing across the system, an intra-cluster and inter-cluster load balancing algorithms are proposed. We have shown that our proposed architecture attains less latency and better throughput with reduced bandwidth usage, through the simulation results.

## About the Authors

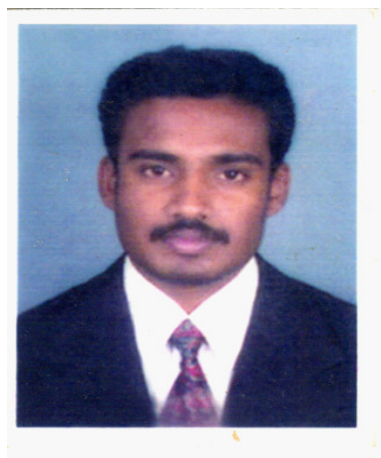

**Mr.S.Ayyasamy** completed his B.E. (Electronics and Communication Engineering) in 1999 from Maharaja Engineering College and M.E. (Computer Science and Engineering) in 2002 from PSG College of Technology, both under Bharathiar University, Coimbatore. Currently he is pursuing PhD degree from Anna University, Coimbatore. He is working as an Assistant Professor, Department of Information Technology at Tamilnadu College of Engineering, Coimbatore. He is a member of various professional bodies like ISTE, CSI and IAENG. His research areas include P2P networks, Overlay Networks, Load Balancing and Quality of Services and having 9 years of teaching experience in Engineering Colleges.

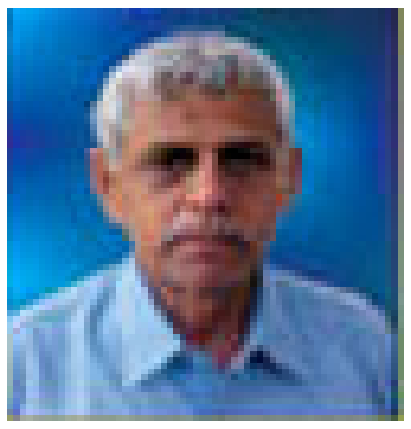

**Dr. S. N. Sivanandam** completed his B.E. (Electrical Engineering) in 1964 from Government College of Technology, Coimbatore, and MSc (Engineering) in Power Systems in the year 1966 from PSG College of Technology, Coimbatore. He acquired PhD in control systems in 1982 from Madras University. He received best teacher award in the year 2001 and **Dhakshina Murthy Award** for teaching excellence from PSG College of technology. He received the citation for best teaching and technical contribution in the year 2002, Government College of Technology, Coimbatore. His research areas include Modeling and Simulation, Neural Networks, Fuzzy Systems and Genetic Algorithm, Pattern Recognition, Multidimensional system analysis, Linear and Non linear control system, Signal and Image processing, Control System, Power System, Numerical methods, Parallel Computing, Data Mining and Database Security. He is a member of various professional bodies like IE (India), ISTE, CSI, ACS and SSI. He is a technical advisor for various reputed industries and engineering institutions.